\documentclass[a4paper]{jpconf}
\usepackage{graphicx}
\usepackage{amsmath}
\usepackage{amsfonts}
\usepackage{amssymb}
\usepackage{textcomp}
\usepackage{slashed}
\begin{document}
\title{$W^*\gamma^*\pi$ form factors in the $\tau^- \to \pi^-\ell^+\ell^- \nu_{\tau}$ decays}

\author{Adolfo Guevara}

\address{Departamento de F\'isica, Centro de Investigaci\'on y de Estudios Avanzados, Avenida IPN 2508, Col. San Pedro Zacatenco, 07360 M\'exico DF, M\'exico}

\ead{aguevara@fis.cinvestav.mx}

\begin{abstract}
 We study the disintegration $\tau^- \to \pi^-\ell^+\ell^-\nu_\tau$, where $l$ denotes either an electron or a muon. This process requires knowledge about the 
$W^*\gamma^*\pi$ interaction, which is modeled using Resonance Chiral Theory. Within this framework we study the sensitivity of the observables of the process, 
such as the branching ratios and the dilepton mass spectra, to the model dependent contributions. The information of the hadronization of the quark currents is 
encoded in form factors which are evaluated using the lightest vector and axial-vector multiplets as degrees of freedom. We determine the couplings by matching the 
asymptotic QCD behavior of the form factors in the limit of an infinite number of colours. The branching ratios we obtain show that the process with $l=e$ should 
be measured soon in $B$ factories.
\end{abstract}

\section{Introduction}
\hspace*{2.5ex}\\
 It is well known that the interaction between quarks and gluons becomes stronger in low-energy processes \cite{PDG}. Due to this non-perturbative characteristic of 
QCD it is advisable to appeal to effective theories. To study this process it is necessary to have precise information about the $W^*\gamma^*\pi$ interaction. 
The relevant form factors have been studied before with Chiral Perturbation Theory ($\chi$PT) \cite{ChPT} in the $\pi\to e^+e^-e\hspace*{1ex}\nu_\e$ decays\cite{Bryman,Scheck,Vaks}, 
where the photon is almost on-shell. The knowledge of this interaction can be used to compute radiative corrections and the couplings needed to calculate  
the lightest pseudoscalar contributions to the hadronic light-by-light piece of the anomalous magnetic moment of the muon\cite{RoigLC}.\\

The difference between the $\tau^- \to \pi^-\ell^+\ell^- \nu_{\tau}$ process and the pion decay is that the process we study may have an energy above the 
chiral symmetry breaking scale, where the $\chi$PT is no longer valid; this is why we employ Resonance Chiral Theory (R$\chi$T)\cite{RChT,Port}. 
R$\chi$T introduces meson resonances as dynamical fields preserving the chiral symmetry properties and carrying out the required calculation 
by means of an expansion in powers of the inverse of the number of colours, $\frac{1}{N_C}$\cite{Nc}. 

\section{Decay amplitude}
\hspace*{2.5ex}\\
 The $\tau^- \to \pi^-\ell^+\ell^- \nu_{\tau}$ \cite{us} decays have not been measured yet, and in order to compute the branching ratios and invariant mass spectra 
we follow the work done in \cite{GuoRoig}, where the process with a real photon is analyzed and the participating vector and axial-vector form factors are 
obtained. As a consequence of the virtuality of the photon, a diagram that vanished in \cite{GuoRoig} now contributes to the amplitude, giving rise to a third 
contributing form factor, $B$ in eq. (\ref{MA}), proportional to the vector two-pion form factor, which was implemented following~\cite{2pi}.

The contributions to the process decay amplitude are the bremsstrahlung off the $\tau$, off the $\pi$ and 
the vector and axial-vector contributions; respectively\cite{us}

\begin{equation}
 i{\cal M}_{IB \tau }=i G_FV_{ud}e^2f_{\pi }\frac{1}{k^2}P_{\mu }l_{\nu }L^{\mu \nu
}\ ,
\end{equation}

\begin{equation}
i{\cal M}_{IB \pi }=i G_FV_{ud}e^2f_{\pi }\frac{1}{k^2}l^{\nu
}\left(\frac{2P_{\nu }(P+k)_{\mu
}}{(P+k)^2-m_{\pi }^2}-\mathit{g}_{\mu \nu }\right)L^{\mu } \ ,
\end{equation}
\begin{equation}\label{MV}
{\cal M}_V=- G_FV_{ud}e^2\frac{1}{k^2} F_V\left(p\cdot k,k^2\right)\varepsilon
_{\mu \nu \rho \sigma }l^{\nu }k^{\rho }p^{\sigma }L^{\mu },
\end{equation}
\begin{eqnarray}\label{MA}
 {\cal M}_A & = & iG_FV_{ud}\frac{2e^2}{k^2}l^{\nu }\left\{F_A\left(p\cdot k,k^2\right)\left[\mathit{g}_{\mu\nu}\left(p\cdot k-k^2\right)-P_{\nu
}k_{\mu }\right]+\right.\nonumber\\
& & \left.B(k^2) k^2\left[\mathit{g}_{\mu \nu }-\frac{(P+k)_\mu P_\nu}{k^2+2P\cdot k}\right]\right\}L^{\mu } \ .
\end{eqnarray}
Where we have defined
\begin{subequations}
 \begin{align}l^{\nu }&=\bar{u}_{\ell^-}\left(P_-\right)\gamma _{\nu }v_{\ell^+}\left(P_+\right), \\ 
  L^{\mu \nu }&=\bar{u}_{\nu _{\tau }}(q)\gamma ^{\mu
  }\left(1-\gamma_5\right)\left(\frac{\gamma \cdot \left(P_{\tau }-k\right)+M_{\tau
  }}{\left(P_{\tau }-k\right){}^2-M_{\tau }{}^2}\right)\gamma ^{\nu
  }u_{\tau }\left(P_{\tau }\right),\\
  L^{\mu }&=\bar{u}_{\nu _{\tau }}(q)\gamma ^{\mu }\left(1-\gamma ^5\right)u_{\tau
  }\left(P_{\tau }\right).
 \end{align}
\end{subequations}

\section{Form factors}
\hspace*{2.5ex}\\
Since the mass of the $\tau$ lepton is larger than the $\chi$PT expansion parameter, the perturbative approach does no longer converge for this process. At this 
energy scale the lightest resonances are used as dynamical fields, because the masses of these resonances are smaller or of the order of the $\tau$ lepton mass. 
This is done with R$\chi$T, using as expansion parameter $\frac{1}{N_C}$ to compute the form factors.\\

Thus, in the framework of R$\chi$T we find that 

\begin{multline}
 F_V(t,k^2)= -\frac{N_C}{24\pi^2F_\pi}+\frac{2\sqrt{2}F_V}{3F_\pi M_V}[(c_2-c_1-c_5)t+(c_5-c_1-c_2-8c_3)m_\pi^2+2(c_6-c_5)k^2]\\
  + \frac{2\sqrt{2}F_V}{3F_\pi M_V}D_\rho(t)[(c_1-c_2-c_5+2c_6)t+(c_5-c_1-c_2-8c_3)m_\pi^2+(c_2-c_1-c_5)k^2]\\
  +\frac{4F_V^2}{3F_\pi}D_\rho(t)[d_3(t+4k^2)+(d_1+8d_2-d_3)m_\pi^2],
\end{multline}
\begin{equation}
 F_A(t,k^2)=\frac{F_V^2}{F_\pi}\left(1-\frac{2G_V}{f_V}\right)D_\rho(k^2)-\frac{F_A^2}{F_\pi}D_{a_1}(t)
        +\frac{F_AF_V}{\sqrt{2}F_\pi}D_{a_1}(t)D_\rho(k^2)(-\lambda''t+\lambda_0m_\pi^2),
\end{equation}
where 
\begin{equation}\label{DR}
 D_R(s) = \frac{1}{M_R^2-s-iM_R\Gamma_R(s)}.
\end{equation}
and
\begin{equation}
 B(k^2) = F_\pi\frac{F_V^{\pi^+\pi^-}|_\rho(k^2)-1}{k^2},
\end{equation}
where $F_V^{\pi^+\pi^-}|_\rho(k^2)$ is the $I=1$ part of the $\pi^+\pi^-$ vector form factor.\\

Given the rather wide widths of the $\rho(770)$ and $a_1(1260)$ resonances, 
their off-shell behavior becomes important and has to be considered in eq.~(\ref{DR}). A consistent and formalism independent definition for the energy-dependent 
spin-one resonance widths was given in \cite{RW}, which we follow for the $\rho(770)$ case. The simplified expression that we use for the $a_1(1260)$ meson was 
obtained in \cite{RW2}.

\section{Branching ratios}
\hspace*{2.5ex}\\
\begin{table}
 \caption{\label{T1} {\footnotesize The central values of the different contributions to the branching ratio of the $\tau^-\to\pi^-\ell^+\ell^-\nu_\tau$ decays are given 
on the left hand side of the table. The error bands due to variations of the couplings are displayed in the right hand side of the table. The error band of the IB-IB 
contributions stems from the uncertainties on the $F_\pi$ decay constant and the $\tau$ lepton lifetime \cite{PDG}.}\\}
\begin{center}
 \begin{tabular}{|c||c|c||c|c|}\hline
  &$l=e$&$l=\mu$&$l=e$&$l=\mu$\\ \hline
IB-IB&$1.461\times10^{-5}$&$1.600\times10^{-7}$&$\pm0.006\times10^{-5}$&$\pm0.007\times10^{-7}$\\
 IB-V&$-2\times10^{-8}$&$1.4\times10^{-8}$&$\left[-1\times10^{-7},1\times10^{-7}\right]$&$\left[-4\times10^{-9},4\times10^{-8}\right]$\\ 
IB-A&$-9\times10^{-7}$&$1.01\times10^{-7}$&$\left[-3\times10^{-6},2\times10^{-6}\right]$&$\left[-2\times10^{-7},6\times10^{-7}\right]$\\
V-V&$1.16\times10^{-6}$&$6.30\times10^{-7}$&$\left[4\times10^{-7},4\times10^{-6}\right]$&$\left[1\times10^{-7},3\times10^{-6}\right]$\\
A-A&$2.20\times10^{-6}$&$1.033\times10^{-6}$&$\left[1\times10^{-6},9\times10^{-6}\right]$&$\left[2\times10^{-7},6\times10^{-6}\right]$\\
V-A&$2\times10^{-10}$&$-5\times10^{-11}$&$\sim10^{-10}$&$\sim10^{-10}$\\ \hline
Total&$1.710\times10^{-5}$&$1.938\times10^{-6}$&$\left(1.7^{+1.1}_{-0.3}\times10^{-5}\right)$&$\left[3\times10^{-7},1\times10^{-5}\right]$\\ \hline
 \end{tabular}
\end{center}
\end{table}

 The form factors are completely determined using short distance constraints \cite{RChT,SD}, which allows us to compute the branching ratio and the $\ell^+\ell^-$ invariant 
 mass spectra. After numerical integration, we obtained the contributions to the total branching ratio of the different parts. These different contributions come 
from the squared moduli of the three amplitudes described in the previous section and their interferences. However, in our approach there is an uncertainty in this 
short distance constraints which is $\lesssim$ 20\%. Taking this into account, the contributions vary as noted in Table \ref{T1}.\\
\begin{figure}
 \begin{center}
 \includegraphics[scale=0.6]{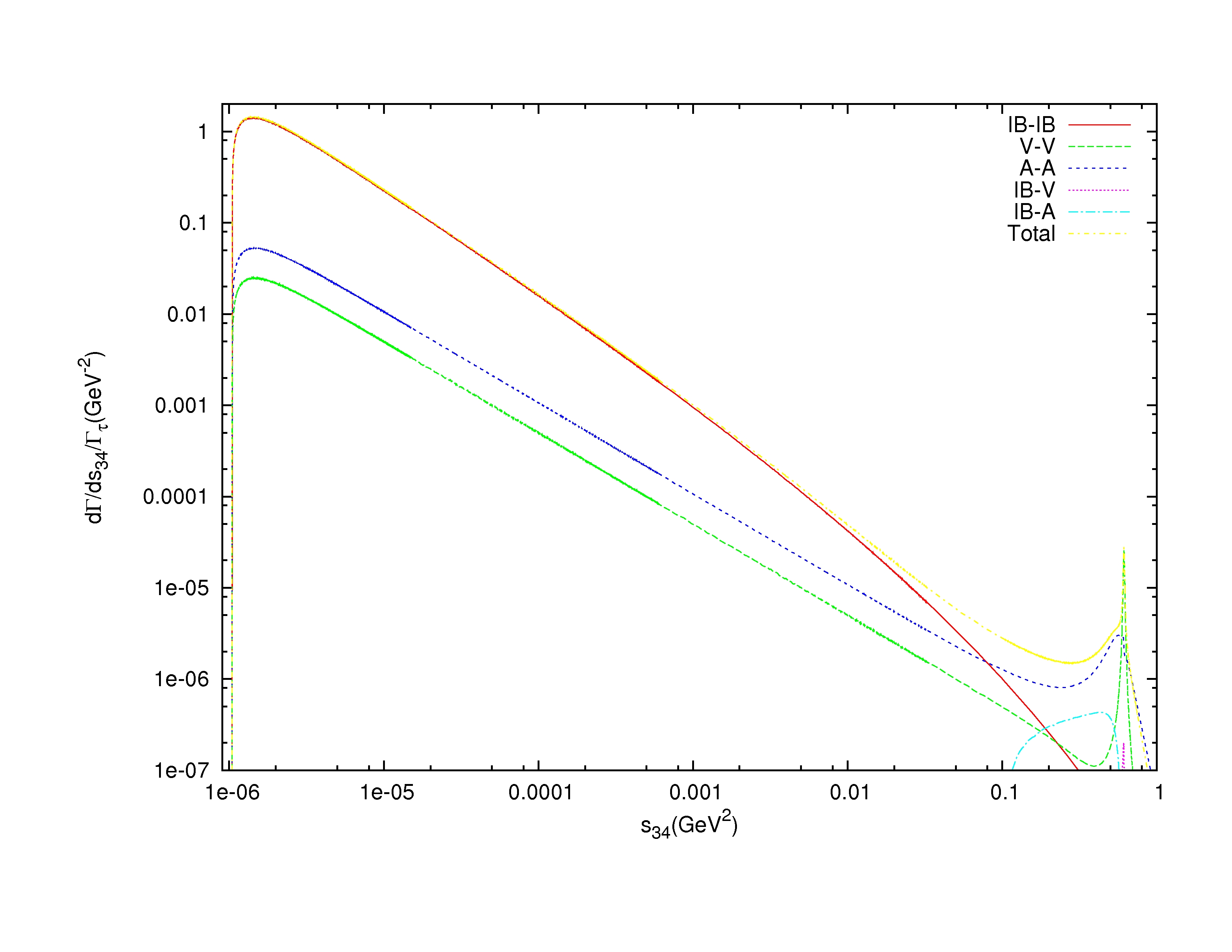}
 \caption{\footnotesize Different contributions to the invariant mass distribution for $\ell=e$.}\label{e}
 \end{center}
\end{figure}
\begin{figure}
 \begin{center}
 \includegraphics[scale=0.6]{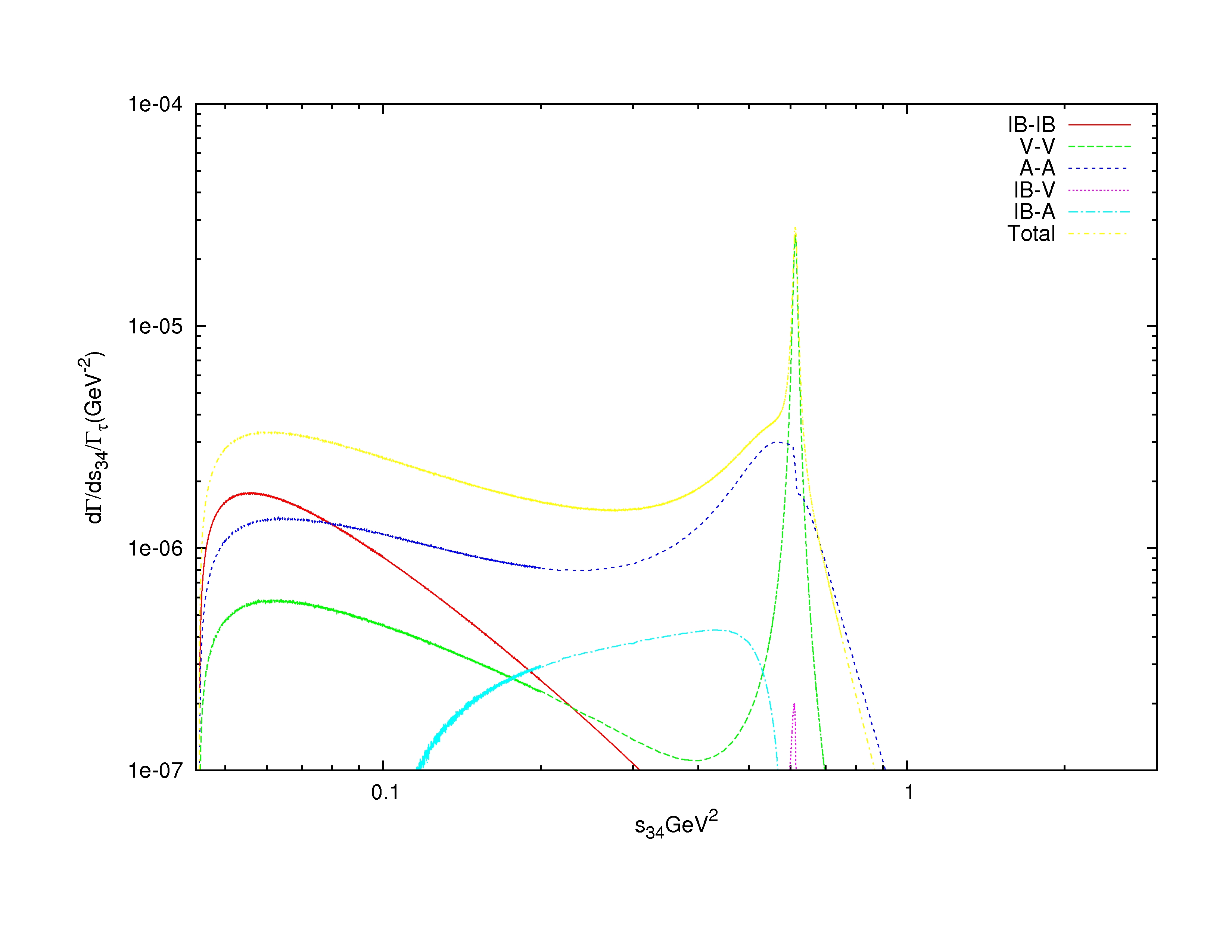}
 \caption{\footnotesize Different contributions to the invariant mass distribution for $\ell=\mu$.}\label{mu}
 \end{center}
\end{figure}

 The normalized invariant-mass distribution of the lepton pair, $$\frac{d\Gamma\left(\tau^-\to\pi^-\ell^+\ell^-\nu_\tau\right)}{ds_{34}}\frac{1}{\Gamma_\tau},$$
 where $\Gamma_\tau$ is the inverse of the $\tau$ lifetime and $s_{34}$ the invariant mass of the $\ell^+\ell^-$ pair, is shown in Figure \ref{e} for $\ell = e$, 
and in Figure \ref{mu} for $l = \mu$. It can be observed in Figure \ref{e} that the structure independent contribution dominates the spectrum for values of 
$s_{34}\lesssim0.1$ GeV$^2$, but for larger values, the axial-vector contribution dominates the rest of the spectrum apart from the $\rho$ peak. The interference 
terms are negative for most of the spectrum and do not appear in the figure because of their smallness. In Figure \ref{mu}, at low invariant mass the A contribution is comparable to the IB 
contribution, but for relatively low values and in the higher energy region the A contribution dominates the spectrum, except around the $\rho$ mass.
The branching fractions shown in Table \ref{T1} were obtained by integrating numerically these invariant-mass distributions and checked from a direct integration 
over the five independent kinematic variables.\\

 The fact that in both cases the structure dependent contribution dominates at $s_{34}\gtrsim0.1$ GeV$^2$ justifies our assumption of including only one multiplet 
of vector and axial-vector resonances. It can be seen in both figures that the A-A contribution is peaked in the $\rho$ mass region, this is due to the  inclusion 
of the $B$ form factor, because it is proportional to the I = 1 component of the electromagnetic di-pion form factor, showing how essential this contribution is. \\
\hspace*{2.5ex}\\
\hspace*{2.5ex}\\
\hspace*{2.5ex}\\
\hspace*{2.5ex}\\

\section{Conclusions}
\hspace*{2.5ex}\\

 We have studied the $\tau^-\to\pi^-\ell^+\ell^-\nu_\tau$ process for the first time, describing in detail the $W^*\gamma^*\pi$ interaction using R$\chi$T. We 
have been able to predict all involved couplings using short-distance QCD constraints. The uncertainty of this procedure has been considered in our error 
estimates.\\

As a result, we predict the branching ratios $BR\left(\tau^-\to\pi^-e^+e^-\nu_\tau\right)=\left(1.7^{+1.1}_{-0.3}\right)\times10^{-5}$ and $BR\left(\tau^-\to\pi^-\mu^+\mu^-\nu_\tau\right)=\left[3\times10^{-7},10^{-5}\right]$, 
where the hadronic error dominates our result for the $\ell=\mu$ case. From these figures, we conclude that the $\ell=e$ decay mode should 
be measured soon in the B-factories. This may also be true for the $\ell=\mu$ case only if the branching fraction happens to be close to 
the upper limit of the range we have given.

\section{Acknowledgements}
\hspace*{2.5ex}

This work has been partially supported by the Spanish grant FPA2011-25948 and Conacyt (M\'exico).

\section*{References}
\hspace*{2.5ex}\\

\end{document}